
\input phyzzx.tex

\def\del{\partial}
\def\nl{\hfill\break}
\def\ni{\noindent}
\def\ket{\rangle}
\def\bra{\langle}
\def\u{\,{\cal U}}
\def\uu{\,{\widehat {\cal U}}}
\def\Duv{[4(uv -\mu/2) \del_u \del_v + 2 (u\del_u + v\del_v) + 1]}
\def\barDuv{4(u \bar u -\mu/2) \del_u \del_{\bar u} + 2 (u\del_u +
\bar u\del_{\bar u}) + 1}
\def\gst{{g_{\rm str}}}
\def\T{{\widehat T}}
\def\delt{\delta T}
\def\ddelt{\delta {\widehat T}}
\def\pphi{{\widehat \phi}}
\def\etab{{\bar \eta}}

\def\R{{\cal R}}
\def\x{{\bf x}}
\def\t{{\bf t}}

\indent \hfill{TIFR-TH-93/05}\break
\indent \hfill April, 1993 \break
\date{}
\titlepage
\title{\bf Wave Propagation in Stringy Black Hole}
\author{Avinash Dhar\foot{adhar@tifrvax.bitnet.},
Gautam Mandal\foot{mandal@tifrvax.bitnet.}
and Spenta R. Wadia
\foot{wadia@tifrvax.bitnet}}
\address{ Tata Institute of Fundamental Research, Homi Bhabha Road, Bombay
400 005, India}
\abstract{We further study the nonperturbative formulation of
two-dimensional black holes. We find a nonlinear differential equation
satisfied by the tachyon in the black hole background. We show that
singularities in the tachyon field configurations are always associated
with divergent semiclassical expansions and are absent in the exact
theory. We also discuss how the Euclidian black hole emerges from an
analytically continued fermion theory that corresponds to the right side
up harmonic oscillator potential.}
\endpage

\REF\DNW{S.R. Das, S. Naik and S.R. Wadia, Mod. Phys. Lett. A4 (1989)
1033; A. Dhar, T. Jayaraman, K.S. Narain and S.R. Wadia, Mod. Phys. Lett.
A5 (1990) 863; S.R. Das, A. Dhar and S.R. Wadia, Mod. Phys. Lett. A5(1990)
799.}
\REF\POLONE{J. Polchinski, Nucl. Phys. B234 (1989) 123.}
\REF\BL{T. Banks and J. Lykken, Nucl. Phys. B331 (1990) 173.}
\REF\BKZ{
E. Brezin, C. Itzykson, G. Parisi and J.B. Zuber, Comm. MAth. Phys. 59
(1978) 35;
E. Brezin, V.A. Kazakov and Al.B. Zamolodchikov, Nucl. Phys. B338 (1990)
673;
D. Gross and N. Miljkovich, Nucl. Phys. B238 (1990) 217;
P. Ginsparg and J. Zinn-Justin, Phys. Lett. B240 (1990) 333;
G. Parisi, Europhys. Lett. 11 (1990) 595.}
\REF\MSW{G. Mandal, A.M. Sengupta and S.R. Wadia, Mod. Phys. Lett. A6
(1991) 1685.}
\REF\WITTEN{E. Witten, Phys. Rev. D44 (1991) 314.}
\REF\SH{For a review see A. Strominger  and J. Harvey, Chicago preprint
EFI-92-41, hep-th/9209055.}
\REF\SW{A.M. Sengupta and S.R. Wadia, Int. J. Mod. Phys. A6 (1991) 1961.}
\REF\GK{D.J. Gross and I. Klebanov, Nucl. Phys. B352 (1990) 671.}
\REF\MOORE{G. Moore, Nucl. Phys. B368 (1992) 557.}
\REF\DJ{S.R. Das and A. Jevicki, Mod. Phys. Lett. A5 (1990) 1639.}
\REF\POLTWO{J. Polchinski, Nucl. Phys. B346 (1990) 253.}
\REF\DMWBH{A. Dhar, G. Mandal and S.R. Wadia, Mod. Phys. Lett. A7 (1992)
3703.}
\REF\MS{E. Martinec and S. Shatashvilli, Nucl. Phys. B368 (1992) 338.}
\REF\SRD{S.R. Das, Mod. Phys. Lett. A8 (1993) 69.}
\REF\RUSSO{J.G. Russo, Phys. Lett. B300 (1993) 336.}
\REF\YONEYA{T. Yoneya, Tokyo (Komaba) preprint UT-KOMABA-92-13,
hep-th/9211079.}
\REF\MV{S. Mukhi and C. Vafa, TIFR and Harvard preprint, TIFR/TH/93-01,
HUTP-93/A002.}
\REF\DMW{A. Dhar, G. Mandal and S.R. Wadia, Mod. Phys. Lett. A7 (1992)
3129.}
\REF\DMWClass{A. Dhar, G. Mandal and S.R. Wadia, Int. J. Mod. Phys.
A8 (1993) 325.}
\REF\GRADSHTEYN{I.S. Gradshteyn and I.M. Ryzhik, {\sl Table of Integrals,
Series and Products}, Academic Press, New York (1965), p. 1039.}
\REF\DDMW{S.R. Das, A. Dhar, G. Mandal and S.R. Wadia, Int. J. Mod. Phys.
A7 (1992) 5165.}
\REF\MSWTWO{G. Mandal, A.M. Sengupta and S.R. Wadia, Mod. Phys. Lett. A6
(1991) 1465.}
\REF\DJR{K. Demeterfi, A. Jevicki and J. Rodrigues, Nucl. Phys. B362
(1991) 173.}

\section{\bf Introduction:}

Two-dimensional string theory can be viewed as one dimensional-matter
(time) coupled to two-dimensional gravity [\DNW,\POLONE,\BL].
Since the latter has a nonperturbative formulation in terms of a
one-dimensional matrix model [\BKZ] (nonrelativistic fermions) we have an
opportunity to study nonperturbative aspects of two-dimensional string
theory. In particular, two-dimensional string theory has a black hole
solution [\MSW,\SW,\SH] and one can begin to explore nonperturbative
aspects of black holes, in particular the important question of the fate
of the classical singularity.

In any attempt to understand the emergence of a non-trivial spacetime in
the matrix model one has to contend with the fact that the
non-relativistic fermions are formulated in a flat spacetime
[\SW,\GK,\MOORE]. However, the ``spacetime'' of the perturbative
collective excitation is a half-plane, the boundary of which is associated
with the classical turning point of the fermions
[\SW,\GK,\DJ,\POLTWO,\MOORE]. On the other hand it is well-known that
the graviton-dilaton system ($G_{\mu\nu}, \Phi$)
for the two-dimensional black hole is
equivalent to a metric $\widetilde G_{\mu\nu} = G_{\mu\nu} \exp(-2\Phi)$
which corresponds to a spacetime that
is flat but has a boundary determined by the condition $\exp(-2\Phi)
\ge 0$. It is this reasoning that makes it plausible that one may be able
to describe a black hole spacetime in the semiclassical limit of the
matrix model.

Recently [\DMWBH] we have
discussed a scalar (tachyon) field theory which in the
semi-classical limit reduces to a scalar field coupled to the
two-dimensional dilaton-black hole.
Other works in this direction include [\MS,\SRD,\RUSSO,\YONEYA,\MV].
References [\MS] and [\MV] deal with the continuum formulation and  relate
the coset black holes with `Liouville' theory. References [\SRD],[\RUSSO]
and [\YONEYA] deal directly with the matrix model.
The  field theory discussed in [\DMWBH] is obtained from
the nonperturbative formulation of the $c=1$ matrix model in terms of the
phase-space distribution operator
$$\uu(p,q,t) = \int_{-\infty}^\infty dx\, \psi^\dagger(q- {x\over 2})
e^{-ipx} \psi(q+ {x\over 2}).$$
Since the correlations of this operator are exactly calculable we are able
to define an exact nonperturbative
quantization of the scalar field. The main result of [\DMWBH] was
that singularities at the position of the black hole singularity ($uv =
\mu/2$ in Kruskal coordinates) that occur in the computation of the
tachyon ground state   do not survive quantization: {\sl the singularities
are only a malady of the semiclassical expansion}.

In the present paper we generalize the last result to show that, even for
generic field configurations, singularities at the centre of the black
hole appear only in semiclassical expansions and like in [\DMWBH] the
exact results are non-singular. We also explore in greater detail the
semiclassical expansion applied to the tachyon field to show that
perturbatively it satisfies a closed non-linear differential equation (the
linear part being the equation of motion for a free field coupled to the
dilaton-black hole). The picture that emerges is the following. In some
regions of space-time where the semiclassical expansion is valid, the
classical physics is described by the above equation of motion for the
tachyon; in those regions where the semiclassical expansion shows
singularity, clearly the differential equation as well as the attendant
notion of spacetime backgrounds are to be discarded in favour of the exact
quantum theory that is defined  in terms of the fermion theory. We
describe the working rules of the quantum theory and explain some unusual
features of the theory associated with non-trivial commutation rules of
the ``tachyon field''. We also demystify the ``hyperbolic transform'' by
exhibiting some of its physically interesting properties and showing that
it is the unique transform which satisfies these properties. Finally we
discuss in brief the Euclidian black hole obtained from an analytically
continued fermion theory.

\section{\bf Quantized Tachyon in Black Hole}

Let us first briefly review some salient points of [\DMWBH].  The basic
construct is the fermion bilinear $\pphi(p,q,t)$ and is defined by a
``hyperbolic transform'' of the quantum phase space density $\uu(p,q,t)$:
$$ \pphi(p,q,t) = \int dp'dq'\, K(p,q| p',q') \uu(p',q',t)
\eqn\twoone$$
where
$$ K(p,q| p',q') \equiv | (p-p')^2 - (q-q')^2 |^{-1/2}
\eqn\twotwo$$
Equation \twoone\ can be regarded as a relation between Heisenberg
operators, where
$$\pphi(p,q,t)= e^{iHt} \pphi(p,q,0) e^{-iHt},\; \uu(p,q,t) = e^{iHt}
\uu(p,q,0) e^{-iHt} \eqn\twothree$$
where
$$H = \int {dp dq\over 2\pi}
h(p,q) \uu(p,q),\qquad h(p,q) = {1\over 2}(p^2-q^2)
\eqn\twofour$$
Clearly equation \twoone\ is also valid for expectation values
$$ \bra \psi|\pphi(p,q,t)|\psi \ket = \int dp'dq'\, K(p,q| p',q')
\bra \psi|\uu(p',q',t) | \psi \ket
\eqn\twofive$$
where $| \psi \ket $ is any state in the Fermi theory.

For the Heisenberg operator $\uu(p,q,t)$ we have the equation of motion
$$ (\del_t + p\del_q + q\del_p)\uu(p,q,t) = 0  \eqn\twosix $$
Used in the definition \twoone, together with \twotwo, this leads to
$$ (\del_t + p\del_q + q\del_p)\pphi(p,q,t) = 0  \eqn\twoseven $$
The last equation implies that if we define the variables
$$ u = e^{-t} (p +q)/2,\; v= e^t (p-q)/2  \eqn\twoeight $$
or equivalently
$$ p = ue^t + ve^{-t},\; q= ue^t - ve^{-t} \eqn\twonine $$
then
$$ \del_t \pphi(ue^t + ve^{-t}, ue^t - ve^{-t},t) = 0 \eqn\twoten$$
This means that we can define
$$\eqalign{
\T(u,v) \equiv & \pphi(ue^t + ve^{-t}, ue^t - ve^{-t},t)  \cr
=&\int du' dv' {\widetilde K}(u,v | u',v')
\uu(u'e^t+v'e^{-t},u'e^t-v'e^{-t},t), \cr}
\eqn\twooneone $$
which is actually independent of $t$. Here
$${\widetilde K}(u,v| u',v') \equiv\, |(u - u') ( v - v')|^{-1/2}
\eqn\twooneonea $$

In [\DMWBH] we observed that if
one considers states $|\psi \ket $ in the
fermion theory such that $ \bra \psi | \uu(p,q,t) | \psi \ket $ differs
from $ \bra \psi_0 | \uu(p,q,t) | \psi_0 \ket $ at most in a small
neighbourhood of the fermi surface $p^2 - q^2 = 2\mu$, then
$$\eqalign{
\delt(u,v) \equiv & \bra \psi | \ddelt(u,v) | \psi \ket
= \bra \psi |\T(u,v) | \psi \ket - T_0(u,v) \cr
\ddelt(u,v) \equiv & \T(u,v) - T_0(u,v) \cr}
\eqn\twoonetwo$$
satisfies
$$ \Duv \delt(u,v) = 0 + o({\delta E \over \mu })  \eqn\twoonethree$$
In the above, $|\psi_0 \ket$ refers to the fermion ground state,
$T_0(u,v)= \bra \psi_0 | \T(u,v)|\psi_0 \ket$ and
$\delta E$ is the maximum spread of energy relative to the fermi surface
in the support of $\bra \psi | \uu(p,q,t) | \psi \ket $.

To this order, therefore, the construct \twoonetwo\ provides us with
classical solutions for a free scalar field coupled to the two-dimensional
black hole. We shall see in the next section that the higher order terms
in $(\delta E / \mu)$ correspond to non-linear terms in $\delt(u,v)$.

Now, as we emphasized in [\DMWBH], expression \twoonetwo\ does more than
just to provide perturbative solutions to \twoonethree. $\delt(u,v)$
is in fact
exactly computable in the fermion theory (in principle for
any state $| \psi \ket$). We will see that typically the exact expression
has a bad perturbative expansion in $(\delta E/ \mu)$ and the divergences
in solutions of \twoonethree\ at the ``black hole singularity'' $uv =
\mu/2$ are a result of such bad expansions. Indeed the exactly computed
$\delt(u,v)$ is free of singularities. In [\DMWBH] we have already
explcitly computed one such example.

We will thus take the attitude that  the fermion theory, through equations
such as \twoonetwo, \underbar{defines} for us quantization of a scalar
field coupled to the two-dimensional black hole. Such a definition is of
course automatically nonperturbative since the fermion theory is so.
In the next few sections we will explore in detail the semiclassical
physics by considering in the fermion theory small fluctuations near the
fermi surface  and show how to use the nonperturbative formalism of the
fermion theory to extract the nonperturbative behaviour.

\section{\bf Properties of the Hyperbolic Transform}

Before launching into properties of the ``tachyon field'' \twooneone\ it
is useful to understand its definition in some more detail.  In this
section we shall mention some remarkable properties of the kernel
\twotwo\ that defines for us the ``hyperbolic transform'' \twoone.
The basic properties of \twotwo\ (equivalently, of \twooneonea) are: \nl
\ni (i) Lorenz covariance: $K(P, Q | P', Q') = K(p,q| p',q')$, where $P =
p\cosh \theta + q\sinh \theta, Q = p\sinh \theta + q \cosh \theta$ and
similarly for $P',Q'$. \nl
\ni (ii) Translational invariance: $K(p,q| p',q') = f(p-p', q-q')$ \nl
\ni (iii) Differential equation:
$$\eqalign{
\,&\Duv \widetilde K(u,v | u',v') =o[u'v'- \mu/2] \cr
\,&\widetilde K(u,v | u',v') = K({u+v\over 2},{u-v\over 2}|{u'+v'\over 2},
{u'-v'\over 2}) \cr}
\eqn\aone$$

The precise form of the last equation is
$$\eqalign{
\Duv & \widetilde K(u,v | u'v') \cr
\,&= (u'v'-\mu/2) \del_u\del_v K(u,v|u',v')\cr}
\eqn\atwo$$
As we shall see, it is equation \aone\ (or \atwo) that is responsible for
the black hole interpretation of the low energy physics of the tachyon.
This is because if one considers a state
$|\psi \ket$ such that the support of
$\bra \psi|\uu(p',q',t) | \psi \ket$ is near the fermi surface, then $\bra
\psi |\T(u,v)  | \psi \ket$ vanishes to leading order according to
\twooneone\ and \aone.

Clearly, property (iii) is very  desirable from the point of view of black
hole physics.  Property (i) implies that the equations of motion are the
same for $\phi(p,q,t)$ and $\u(p,q,t)$; this implies that the reduced
variables $u,v$ used in both cases have the same physical interpretation.
Property (ii) is directly related to the fact that the hyperbolic
transform becomes local in Fourier space; in other words, the (double)
Fourier transform of $\phi(p,q,t)$'s ($\widetilde \phi(\alpha,
\beta, t)$'s) are basically rescalings of $W(\alpha,\beta,t)$'s [\DMW,
\DMWBH] which ensure that the $\widetilde \phi$'s have an algebra that has
a classical limit. The latter property implies that the classical action
written in terms of $\phi(p,q,t)$ has an $\hbar\to 0$
limit which, once again is rather crucial.

We now prove that \twotwo\ is the {\bf unique kernel} satisfying all the
three properties mentioned above.

\ni Proof: \nl
\ni Properties (i) and (ii) imply that $K$ is some function of only the
combination $(p-p')^2 - (q-q')^2$:
$$
K(p,q | p', q') = g((p-p')^2- (q-q')^2)) \longleftrightarrow
\widetilde K(u,v | u',v') =  f(x),\;
x= (u-u')(v-v')
\eqn\athree
$$
Property (iii) states that if we choose
$u' = q_0 e^\theta/2, v'= -q_0 e^{-\theta}/2, q_0\equiv
\sqrt{-2\mu}$ so that
$u'v' = \mu/2$ (note that $\mu$, the fermi energy, is negative in our
convention), then
$$\Duv \widetilde K(u,v | q_0e^\theta/2, -q_0e^{-\theta}/2) = 0
\eqn\afour$$
For this choice of $u',v'$ we have $x = uv -\mu/2 + [q_0/2] (ue^{-\theta}
- ve^\theta)$. Using \athree\ for $\widetilde K$, and introducing the
notation $y = uv - \mu/2$ , we get from \afour
$$ 2xf'(x) + f(x) + y[ 6f'(x) + 4xf''(x)]= 0
\eqn\afive$$
This equation must be identically satisfied for all $y$, which implies
that both the $y$-independent term and the coefficient  of $y$ must vanish
in \afive. Curiously the first condition implies the second condition, so
we need not separately consider the second condition. Thus we get
$$ 2xf'(x) + f(x) = 0
\eqn\asix$$
The above equation is solved by
$$ f(x) = \hbox{constant}\; |x|^{-1/2}
\eqn\aseven$$
which proves that (modulo an overall constant)
$$ K(p,q|p',q') = |(p-p')^2 -(q-q')^2|^{-1/2}
\longleftrightarrow \widetilde K(u,v| u',v') = |(u-u')(v-v')|^{-1/2}$$

\section{\bf Non-linear Differential Equation for the Tachyon}

In this section we discuss the semiclassical physics of the tachyon
$\T(u,v)$ in detail and derive a closed non-linear differential equation
for $\delt(u,v) = \bra \psi | \T(u,v) | \psi \ket - T_0(u,v)$ in a
semiclassical expansion.

Let  us consider states $| \psi \ket $ which satisfy
$$ \bra \psi | \uu(p,q,t) | \psi \ket = \vartheta ([ p_+(q,t)- p] [p -
p_-(q,t)]) + o(\hbar).  \eqn\threeone $$
where $\vartheta(x) = 1$ if $x>0$ and $=0$ otherwise.
This corresponds to the ``quadratic profile'' ansatz [\POLTWO,\DMWClass].
We recall that
in the $\hbar \to 0$ limit, the classical $\u(p,q,t)$ (expectation value
in a state) satisfies $\u^2(p,q,t) = \u(p,q,t)$ which implies that
$\u(p,q,t)$ is the
characteristic function for some region; the quadratic profile ansatz
assumes that there is one connected region with boundary given by
$ (p- p_+(q,t) ) ( p - p_- (q,t)) = 0. $
As is well-known, in the ground state $| \psi_0 \ket$, the boundary is the
fermi surface itself: $ p^2 - q^2 - 2\mu=0,$ which correspond to
$$ p_\pm^0 (q,t) = p_\pm^0 (q) = \pm \sqrt{q^2 + 2\mu}\;\vartheta (q^2 +
2\mu).
\eqn\threetwo $$
We shall use the notation
$$ p_\pm(q,t ) \equiv p_\pm^0 (q) + \eta_\pm(q,t)  \eqn\threethree$$
Let us compute  $\delt(u,v)$ in the state \threeone
(ignoring the $o(\hbar)$ terms for the moment), using \twoonetwo. We get
$$\eqalign{
\delt(u,v)=& \int_{-\infty}^\infty dq\int^{\eta_+(q,t)}_0 {
dp' \over |[2ue^t-(p'+ p^0_+(q)+q)][2ve^{-t}- (p'+p^0_-(q)-q)]|^{1/2} }
\cr
-& \int_{-\infty}^\infty dq\int^{\eta_-(q,t)}_0 {
dp' \over |[2ue^t-(p'+ p^0_-(q)+q)][2ve^{-t}- (p'+p^0_-(q)-q)]|^{1/2} }
\cr }
\eqn\threefour$$
Let us now make some further assupmtions about the state $|\psi\ket$,
namely that the fluctuations $\eta_\pm(q,t)$ are non-zero only for $q <
-q_0 \equiv - \sqrt{-2\mu}$ and that in this region $|\eta_\pm(q,t)| <<
|p^0_\pm(q)|$. In other words, we are considering ``small'' fluctuations
near the left branch of the fermi surface in the classically allowed
region.
Under these assumptions we can expand the integrand in a Taylor series in
$p'$ around $p'=0$. The $p'$-integrals now are easy to do, giving powers
of $\eta_\pm(q)$. The $q$-integrals that remain are now effectively
between $-\infty$ and $-q_0$ [$q_0= \sqrt{-2\mu}$], so one can use a
different integration variable $\tau$, the ``time of flight'', given by
$$ q = -q_0 \cosh \tau, \; p^0_\pm (q) = \pm q_0 \sinh \tau
\eqn\threefoura
$$
If we also define the  rescaled  functions $\etab_\pm(\tau, t) = | p_\pm^0
(q) | \eta_\pm (q,t) $, we ultimately get
$$\eqalign{
\delt&(u,v) ={1\over 2}
\int_0^\infty d\tau \Big[k_+(u,v|\tau,t) \etab_+(\tau,t) -
k_-(u,v | \tau, t)\etab_-(\tau,t)] \cr
- {1\over 8}&
\int_0^\infty {d\tau\over p^0_+(q)}(e^{-t}\del_u +e^t
\del_v)k_+(u,v|\tau,t)
\etab_+(\tau,t)^2  -
(e^{-t}\del_u+e^t
\del_v)k_-(u,v | \tau, t)\etab_-(\tau,t)^2 \Big] +o(\etab_\pm^3)\cr}
\eqn\threefive$$
where
$$k_\pm(u,v| \tau,t) \equiv |(ue^t + {q_0\over 2}
e^{\mp \tau})(ve^{-t}- {q_0\over 2}e^{\pm
\tau })|^{-1/2}, \quad q_0 \equiv \sqrt{-2\mu}
\eqn\threefivea$$

{\bf Relations between $\etab_\pm$ and $\delt(u,v)$:}

Equation \threefive\ is important in that it builds a correspondence
between the semiclassical quantities of the fermion theory and those of
the $\delt(u,v)$-theory. To understand it better, let us first choose a
different coordinatization for the $u,v$-space. Let us define\foot{
We use boldface letters so as to distiguish $\t$ from the time $t$ of the
fermion theory.}
$$ \x = {1\over 2} \ln |{uv\over \mu/2}|, \quad \t = {1\over 2}\ln |v/u|
\eqn\threefiveb$$
This is not a one-to-one map. Let us consider for the moment the quadrant
of the $u,v$-space where $u>0, v>0$. In that case we can write down the
inverse maps as follows:
$$ u = {q_0\over 2}e^{\x - \t},\quad v={q_0\over 2}e^{\x +\t}
\eqn\threefivec$$
Now recall that by definition of $\delt(u,v)$ ({\it cf.} the remark
about the $t$-independence of the right hand side of \twooneone), if
$\etab_\pm (\tau, t)$ satisfy their equations of motion [these can be
derived by tracing their definition back to $\u(p,q,t)$ and they read as
$$ (\del_t - \del_\tau)\etab_\pm= {1\over 2} \del_\tau \big(\etab_\pm^2
\over [p^0_+]^2 \big)
\eqn\threefived]$$
then the right hand side of \threefive\ is actually $t$-independent. This
means that we can choose $t$ to be anything we like. It is most useful to
choose
$$ t = \t
\eqn\threefivee$$
on the right hand side of \threefive. This equation now reads,
to leading order, as
$$ \delt(\x,\t) = {1\over 2}
\int_0^\infty d\tau [\widetilde k_+(\x |\tau) \etab_+(\tau,
\t) -\widetilde k_-(\x |\tau) \etab_-(\tau, \t)] + o(\etab_\pm^2)
\eqn\threefivef $$
where
$$ \widetilde k_\pm (\x,\tau) \equiv {2\over q_0}
|(e^\x + e^{\mp\tau}) (
e^\x - e^{\pm\tau)}|^{-1/2}
\eqn\threefiveg $$
Note that for large $\x$,
$${q_0\over 2}\widetilde k_-(\x,\tau)\exp[\x] = |1 +
\exp(\tau - \x)|^{-1/2}+ o(e^{-\x})$$
The first term is  similar to a low-temperature
Fermi-Dirac distribution (the fact that the power is $-1/2$ instead of
$-1$ does not materially affect the arguments). In fact,
one can show that the
for very large $\x$ it behaves like $ \vartheta(\tau -\x)$ and
corrections to it are like increasing powers of $\del_\x$ on
the $\vartheta$-function. Similar expansions are also available for
$k_+(\x,\tau)$. The precise statements for these $\del_\x$-expansions are
the following:
$${\cal T}(\x,\t) = {1\over 2} \{{\cal D}_+\etab_+ - {\cal
D}_-\etab_- \} + o(e^{-\x} \etab_\pm) + o(\etab_\pm^2)
\eqn\threefiveh$$
where
$$ {\cal T}(\x,\t) \equiv |uv|^{1/2} \delt(u,v) \eqn\threefiveha$$
$$ {\cal D}_\pm \equiv I_\pm (\del_\x)+ I_\pm (1/2 - \del_\x)
\eqn\threefivehaa$$
$$ I_\pm (\alpha) = \alpha^{-1} \;_2F_1({1\over 2}, \alpha; \alpha +1;
\mp 1) $$
where $\;_2F_1$ is the standard Hypergeometric function [\GRADSHTEYN].
Using its properties one can write down an expansion for ${\cal D}_\pm$ in
$\del_\x$. The expansion begins with $\del_\x^{-1}$. Defining
$$ \eta_\pm = \pi_\eta \pm \del_\tau \eta,
\eqn\threefiveha$$
where $\eta$ is the ``tachyon'' field that is associated with the standard
$c=1$ matrix model and $\pi_\eta$ is its conjugate momentum, we can write
down a derivative expansion for ${\cal T}(\x,\t)$:
$$ {\cal T}(\x,\t) = \eta(\x,\t) + o(\del_\x\eta, \pi_\eta).
\eqn\threefivehb$$
The identification of the ``black hole tachyon'' field with the standard
$c=1$ tachyon field in the asymptotic ($\x \to \infty$) is rather
remarkable. As a result of this $n$-point functions of ${\cal T}(\x,\t)$
are the same as those of the $c=1$ tachyon at extreme low energy.

Finally, relation \threefiveh\ can be inverted to give $\etab_\pm$
in terms of
${\cal T}$:
$$ \etab_\pm = ({\cal D}_\pm \del_\x)^{-1} \del_\pm {\cal T}(\x,\t)+
o(e^{-x} {\cal T}) + o({\cal T}^2), \quad
\del_\pm = \del_\t \pm \del_\x
\eqn\threefivei$$
We will use this relation below to obtain a nonlinear differential
equation for ${\cal T}(\x,\t)$.

We wish to emphasize
that the above analysis can be repeated in other coordinates which are
valid all through the Kruskal diagram (for instance in the light cone
coordinates themselves) However, the formulae look more complicated.

{\bf Differential Equation:}

Let us now go back to the other consequences of Eq. \threefive.
Eq. \afour\ of Sec. 3 ensures that
$$ \Duv k_\pm (u,v | \tau, t) =0 $$
Using this, and applying the above differential opertor to \threefive\ we
get
$$\eqalign {\Duv& \delt(u,v) \cr
={1\over 2}
&\del_u\del_v \int_0^\infty d\tau[k_+ \etab_+^2 + k_-\etab_-^2] +
o(\etab_\pm^3) \cr}
\eqn\threesix$$
Note that the linear term dropped out because of the special properties of
the kernel.
Using the $\x,\t$-coordinatization, \threesix\ can be written as
$$ D_{\x, \t} {\cal T}(\x, \t) = - {e^{-2\x}\over |\mu|} [
e^{ \x/2} \del_\x\{ e^{-\x/2}({\cal D}_+\etab_+^2 + {\cal D}_-\etab_-^2
)\}] + o(e^{-3\x}\etab_\pm^2) + o(\etab_\pm^3)
\eqn\threeseven$$
where
$$\eqalign{
D_{\x,\t} =& e^\x \Duv e^{-\x} \cr
=& (1+ e^{-2x})(\del_\x^2 - \del_\t^2)+  e^{-2\x}(2\del_\x+1)\cr}
\eqn\threeeight $$
Finally, by using \threefivei\ to convert the $\etab_\pm$ back into ${\cal
T}$, we get the following closed differential equation  in ${\cal T}$ upto
quadratic order:
$$\eqalign{D_{\x, \t} {\cal T}(\x, \t) = - {e^{-2\x}\over |\mu|} [
e^{ \x/2} \del_\x\{ e^{-\x/2}({\cal D}_+[({\cal
D}_+\del_\x)^{-1}\del_+{\cal T} ]^2 +&
{\cal D}_-[({\cal
D}_-\del_\x)^{-1}\del_-{\cal T} ]^2
)\}]\cr
\;&+ o({\cal T}^3) + o(e^{-3x}{\cal T}^2)\cr}
\eqn\threenine$$

\section{\bf Exact Quantum Theory Does Not See Black Hole Singuarity:}

In this section we will analyze the nature of singularities that occur in
$\delt(u,v)$ at $uv = \mu/2$ (which is the position of the curvature
singularity of the black hole metric $ds^2= dudv/( uv - \mu/2)$). We will
see that these singularities occur as a result of making badly divergent
semiclassical expansions and they are not present in $
\delt(u,v) $ when calculated exactly.

Let us consider a state $| \psi \ket $ in the fermion theory which, like
in the previous section, represents fluctuations in the neighbourhood of
the left branch of the  fermi surface $p^2 - q^2 = 2\mu$ (generalizations
are obvious). In this region the fermion phase space can be coordinatized
by $(p,q) = (R \sinh \theta, -R\cosh \theta), R>0.$ As explained earlier,
the support of $\bra \psi | \uu(p,q,t) | \psi \ket $ in the limit $\hbar
\to 0$ defines a region $\R(t)$ occupied by the fermi fluid at time $t$.
For the ground state $| \psi_0 \ket$  this region is given by $\R_0 = \{
\infty > R \ge R_0\equiv\sqrt{-2\mu}, \infty > \theta > -\infty \}$ (plus
its mirror image on the right half of the phase plane). For simplicity of
calculation, let us choose for the moment a state $| \psi \ket $ so that
the region $\R(t)$ has a particularly simple geometry. To be specific, we
choose that $\R(t=0)$ is obtained from $\R_0$ by adding a region $\delta
\R = \{ R_0 \ge R \ge R_1, \theta_1 \ge \theta \ge
\theta_2 \}$ and subtracting a
region $\widetilde{\delta\R} = = \{ \widetilde R_1 \ge R \ge
R_0, \widetilde
\theta_1 \ge \theta \ge \widetilde \theta_2 \}$.
We will call $\delta \R$ the
``blip'' and $\widetilde {\delta \R}$ the ``antiblip''; basically the
state $| \psi \ket $ is created from the ground state $| \psi_0 \ket $ by
removing fermions from the region $\widetilde {\delta \R} \subset \R_0$
and placing them in the region $\delta \R$ just outside the filled fermi
sea. Fermion number conservation is achieved by choosing the areas of
$\delta \R$ and $\widetilde {\delta \R}$ to be the same, which is
equivalent to the condition that
$$ (\theta_1 - \theta_2) (R_1^2 - R_0^2) = (\widetilde{\theta_1} -
\widetilde {\theta_2}) (R_0^2 - \widetilde R_1^2) $$
The region $\R(t)$ at non-zero times $t$ is simply obtained by shifting
the $\theta$-boundaries of both the blip and the antiblip by $t$. Using
this, one can easily write down the expression for $\delt(u,v)$ for this
state:
$$ \delt(u,v) = \delt_b(u,v) + {\widetilde\delt}_b (u,v) + o(\hbar)
\eqn\fourone
$$
where
$$ \delt_b (u,v) = {1\over 2}
\int^{R_0}_{R_1} RdR \int^{\theta_2}_{\theta_1} d\theta
| ( u + Re^{-\theta } ) (v - Re^\theta )|^{-1/2}
\eqn\fourtwo
$$
represents contribution of the `blip'.
The contribution of the `antiblip',
$\widetilde{\delt}_b(u,v),$ is a similar expression involving the
tilde variables. It is important to note that there are
$\hbar$-corrections to
\fourone, as $\u(p,q,t)$ is actually a characteristic function plus
$o(\hbar)$ terms.

A remark is in order about two seemingly different expansions that we are
making in this paper. One is in $\hbar$, and the other is in $|\delta
E/\mu|$. Ultimately, as it turns out, both are expansions in $\gst$. For
the moment, in \fourone\ we have made an explicit $\hbar$-expansion, and no
$|\delta E/\mu|$ expansion yet. What we will show in the present
example is that
it is this latter expansion that is badly divergent and results in
increasingly singular behaviour at $uv = \mu/2$, and if one treats
expressions like \fourtwo\ without a $|\delta E/\mu|$ expansion, then
$\delt(u,v)$ does not have any singularities at $uv = \mu/2$. More
generally, we will argue that singularities are invariably absent whenever
one performs the $|\delta E/\mu|$ resummation; the $\hbar$-corrections
coming from corrections like those present in \fourone\ do not affect the
conclusions {\it vis-a-vis} singularities.

Let us analyze the singularities of \fourtwo\ (treatment of $\widetilde
\delt_b
(u,v)$ is similar). Clearly singularities can arise only when the
expression inside the square root vanishes. It is also clear that a linear
zero inside the square root is not a singularity (recall that $\int dx\,
x^{-1/2}$ is not singular), we must have a quadratic zero. In other words,
both factors must vanish. This can happen only if $u<0, v>0$. Let us
choose the parametrization $u = -r e^{-\chi}/2, v= re^\chi/2$. We get
$$\delt_b(u,v) ={1\over 2}
\int^{R_0}_{R_1} RdR \int^{\theta_2}_{\theta_1} d\theta | (R-r)^2 -
4Rr\sinh^2({\theta - \chi
\over 2})|^{-1/2}
\eqn\fourthree$$
If we look at any one of the integrals separately, over $\theta$ or $R$,
we see a logarithmic singularity at $R=r,\theta= \chi$ provided this point
is included in the range of integration. Let us do the $\theta$ integral
first. If $r$ is outside the range $[R_1, R_0]$ we do not have any
singularities. If $r\in (R_1, R_0)$, and $\chi\in (\phi_1, \phi_2)$,
it is easy to see  that for $R \approx r$
the $\theta$ integral behaves as $\ln | R - r|$. Let us assume
that the range $[R_0, R_1]$ is small (compared to $R_0$, say, which means
that the blip consists of small energy fluctuations compared to the fermi
energy) so that $R\approx r$ through the range of integration (this is
only a simplifying assumption and the conclusions do not depend on it). We
get (for $\phi_2 > 1/2\ln(-v/u) > \phi_1$)
$$\eqalign{
\delt_b(u,v) \sim& \int_{R_1}^{R_0} RdR \ln |R -r| \cr
=& (R_0 - r)\ln |R_0 - r| - (R_1 - r) \ln |R_1 - r| - (R_1-R_0)\cr
\approx & (-\mu/2)^{-1/2}[ (uv - \mu/2)\ln| uv - \mu/2| - (uv - \mu/2 -
\Delta/2) \ln | uv - \mu/2 - \Delta/2| \cr}
\eqn\fourfour$$
In the last line, we have put in the values $R_0 = \sqrt{-2\mu}, r =
\sqrt{ -4uv}$ and defined $R_1 \equiv \sqrt{2(-\mu - \Delta)}, \Delta> 0$
($\Delta $ thus measures the maximum energy fluctuation of the blip from
the fermi surface).  Since $R_0 \approx r \approx R_1$ we have used
$R_{0,1} - r \approx (R_{0,1}^2 - r^2)/(2R_0)$ and also $R_0 - R_1
\approx (R_0^2 - R_1^2)/(2R_0)$.

It is easy to see that $ \delt_b(u,v)$ has no singularities\foot{
This statement is true all over $u,v$-space including the horizon.
}. However it is
also easy to see that it develops singurities as soon as one
attempts a semiclassical expansion in $\Delta/\mu$; to be precise one gets
$$\delt_b(u,v) \sim |\Delta/\mu| \ln| uv - \mu/2| + |\Delta/\mu|^2 (uv -
\mu/2)^{-1} + \cdots \eqn\fourfive$$
where once again $1/2 \ln (-v/u) \in (\phi_1, \phi_2)$ (for $1/2 \ln
(-v/u)$ outside this range there are no singularities).

Thus, we see that the tachyon solution \fourtwo\ develops a singularity at
$uv= \mu/2$ at the level of a $\Delta/\mu$ expansion, though the full
solution does not.

What does the above example teach us for the general scenario?
In general, the tachyon solution would be given by expressions like
$$ \delt(u,v) ={1\over 2}\int RdR \int d\phi f(R,\theta) |
4 ( u + Re^{-\theta}) ( v - R e^\theta)|^{-1/2} \eqn\foursix$$
where $f(R,\theta) \equiv \delta\u(R\sinh\theta, -R\cosh\theta)$ and  the
only thing that we have assumed is that the support of $\delta\u
\equiv \bra \psi |\uu | \psi \ket - \bra \psi_0| \uu | \psi_0 \ket $ is
confined to the region $p^2 < q^2$ (the generalization to the other
region is straightforward). Note that the expression \fourtwo\ can be
recovered by putting
$$\delta\u = \vartheta [(R - R_0)(R_1 - R)] \vartheta [(\theta - \theta_1)
(
\theta_2 - \theta)] \eqn\fourseven $$
where $\vartheta(x)= 1$ if $x>0$ and $0$ otherwise.
The integral \foursix\ consists of contributions from $R>0$ (left half of
the phase plane) and from $R<0$. Let us look at the $R>0$ part first. Once
again the singularities can only come from $u<0, v>0$ region of the $u,v$
space. Using the same parametrization  as in \fourthree\ we get
$$\delt(u,v) ={1\over 2} \int RdR \int d\theta f(R,\theta) | (R-r)^2 -
4Rr\sinh^2({\theta - \chi\over 2})|^{-1/2}
\eqn\foureight$$
The basic lesson of the previous example is that even though each
one-dimensional integral, taken separately over $R$ or $\theta$, has a
logarithmic singularity if the point $R=r, \theta=\chi$ is included in the
support of $f(R, \theta)$, the second integral smooths out that
singularity. (In the previous example we did the $\theta$ integral first
to find logarithmic singularity and the $R$-integration smoothed that out;
it could as easily have been done the other way around.) In fact the issue
is that of a {\bf two-dimensional} integration of the sort $\int dxdy\,
f(x,y)| x^2 - y^2 |^{-1/2}.$ For a smooth function $f(x,y)$ the possible
singularity at $x=y=0$ (a linear zero in the denominator) is washed away
by a stronger (quadratic) zero in the integration measure.
A singularity can be
sustained only if $f(x,y)$ has a pole or a stronger singularity at
$x=y=0$. However, in our case the quantum phase space density $\u(p,q,t)$
cannot have such singularities. The reason is that
the fermion field theory states
that we are concerned with are $W_\infty$-rotations of the ground state,
and therefore the corresponding $\u(p,q,t)$ is also a $W_\infty$-rotation
of the ground state density $\u_0(p,q)$. Since the latter is a smooth
distribution and a unitary rotation cannot induce  singularities,
we see that $\u(p,q,t)$ must be non-singular. The tachyon
field configurations
constructed by integrals such as \foureight\ are therefore
non-singular too. Another way to
think of this is to use a two-step argument: the $\hbar\to 0$ limit of
$\u(p,q,t)$ is obtainable by a classical area-preserving diffeomporphism
(element of $w_\infty$) and is certainly nonsingular, being given by the
characteristic function of a region equal to the $w_\infty$-transformed
Fermi sea. This implies that the tachyon field constructed from it is
already non-singular; incorporation of $\hbar$-effects further smoothen
these disctributions.

We conclude therefore that the exact quantum theory does not permit any
singlarities in the tachyon field configuration.

\section{\bf Some Novel Features of the Quantum Field Theory of $\T(u,v)$}

So far we have discussed the semiclassical physics of $\delt(u,v)$
including its low energy differential equation  and in
the last section we have seen how to compute the expectation values of
$\ddelt(u,v)$ beyond the semiclassical expansion using the fermion theory.
In this section we discuss in more detail some novel features of
the two-dimensional quantum field theory of $\ddelt(u,v)$ defined by the
underlying fermion theory.

Let us first discuss how $\ddelt(u,v) \equiv \T(u,v) - T_0(u,v)$ which is
{\it a priori} defined in terms of an on-shell three-dimensional field
$\uu(p,q,t)$ defines a Heisenberg operator in a {\sl two-dimensional}
field theory. Basically we use the fact that the right hand side of
\twooneone\ is actually independent of $t$ to put $t$ equal to the
``time'' of the $u,v$-space. To fix ideas, let us consider the
coordinatization \threefiveb-\threefivec\ of the $u,v$-space where $\x,\t$
correspond to space and time. This of course limits the discussion to the
quadrant $\{u>0, v>0\}$ but the discussion holds just as well
for more global choices
of ``time'' coordinates also. Now, in these coordinates,
writing $\ddelt(u(\x,\t), v(\x,\t))$ as $\ddelt(\x,\t)$ by an abuse of
notation, we have
$$\eqalign{
\ddelt(\x,\t)& \; \cr
= \int dudv&\;
| (e^x-ue^\t)(e^x-ve^{-t})|^{-1/2}\delta \uu(ue^t + ve^{-t}, ue^t -
ve^{-t}, t) \cr
= \int dudv&\;
| (e^x-u)(e^x-v)|^{-1/2}\delta \uu(ue^{t-\t}+ ve^{\t-t},
ue^{t-\t}-ve^{\t-t},
t) \cr}
\eqn\sixone$$
Here
$$ \delta \uu(p,q,t) \equiv \uu(p,q,t) - \bra \psi_0 | \uu(p,q,t) | \psi_0
\ket $$
Since
these expressions are actually independent of $t$, we can choose the
``gauge''
$$ t = \t
\eqn\sixtwo$$
which gives
$$\ddelt(\x,\t) = \int dudv\; |(e^x - u)(e^x -v)|^{-1/2}
\delta \uu(u+v,u-v, \t) \eqn\sixthree$$
Note that the fields on both sides of \sixthree\ are evaluated at the same
time $\t$. More explcitly we see that
$$\ddelt(\x,\t) = e^{iH\t} \ddelt(\x,0) e^{-iH\t} \eqn\sixfour$$
where
$$\ddelt(\x,0) = \int dudv\; |(e^x - u)(e^x -v)|^{-1/2}\delta \uu(u+v,u-v,
0)
\eqn\sixfive$$
The hamiltonian is the same as in \twofour. Eq. \sixfour\ tells us that
$\ddelt(\x,\t)$ is a Heisenberg operator in a two-dimensional field
theory. Eq. \sixthree\ allows us to write down time-ordered products of
$\ddelt(\x,\t)$'s in terms of time-ordered products of the $\delta
\uu(p,q,t)$'s:
thus
$$\eqalign{
\bra &\psi| {\cal T}(\ddelt(\x_1,\t_1)\cdots \ddelt(x_n,\t_n)) | \psi
\ket \cr
=&\int du_1dv_1\cdots du_ndv_n |(e^{x_1}-u_1)(e^{x_1}-v_1)|^{-1/2} \cdots
|(e^{x_n}- u_n)(e^{x_n} - v_n)|^{-1/2}\times \cr
\qquad &{\cal T}(
\delta \uu(u_1+v_1, u_1-v_1,\t_1)\cdots \delta \uu(u_n + v_n, u_n - v_n,
t_n))
\cr}
\eqn\sixsix$$
In this way the expectation values of time-ordered products of
$\T(u,v)$ can be computed from the fermion theory. As we have stressed
earlier, in principle these contain answers  to all dynamical
questions in the theory. However, these correlation functions are not
related  to usual particle-scattering amplitudes in the standard fashion,
parimarily because:

$\bullet$ $[\ddelt(\x_1,\t), \ddelt(\x_2, \t)] \ne 0$, {\it i.e.},
the field $\ddelt(\x, \t)$ does not commute with itself at equal times.

In fact the non-trivial commutation relation is a direct consequence of
the $W_{\infty}$ algebra.  As we should expect, the field $\T(u,v)$ bears
a close resemblance to the spin operator in a magnetic field. In both
cases the symplectic structures (ETCR in the quantum theory) are
non-trivial. We know that in case of the spin, dynamical questions are
better formulated in terms of coherent states $| {\bf n}\ket$ satisfying
$\bra {\bf n} |\widehat{\bf S} | {\bf n} \ket = {\bf n}$. Questions such
as how $| {\bf n} \ket$ evolves in time are equivalent to calculating the
dynamical trajectory $\bra \widehat {\bf S(t)} \ket $. In the present case
$\delt(u,v) \equiv \bra \ddelt(u,v) \ket$
plays a role exactly similar to this object.

$\bullet$ Two-point function $\ne $ ``Propagator''.

We have seen in Sec. 4 that $\delt(u,v)$, or equivalently ${\cal
T}(\x,\t)$, satisfies the `black hole' differential equation \threenine.
In an ordinary scalar field theory such a thing would imply
$$ D_{\x,\t} G_2(\x,\t| \x',\t') = (1+ e^{-2\x})
\delta(\x-\x') \delta (\t-\t') \eqn\sixseven $$
where
$$
G_2(\x,\t|\x',\t') \equiv \bra \psi_0 | T(
{\widehat{\cal T}}(\x,\t) {\widehat{\cal T}} (\x',\t') | \psi_0 \ket
\eqn\sixeight$$
The prefactor $1 + \exp(-2\x)$ is equal to  $1/\sqrt
{\hbox{det}\, g}$ in $(\x,\t)$ coordinates, required to make the
$\delta$-function covariant. The operator
${\widehat{\cal T}}$ is defined as
$$\widehat{\cal T}(\x, \t) \equiv | uv |^{-1/2} \ddelt(u,v).
\eqn\sixeighta$$
The symbol
$T$ in \sixeight\ denotes time-ordering in the time $\t$.

In our case, \sixseven\ is not true because of the non-standard
commutation relations of the $\widehat{ \cal T}$ field. It is easy to
derive that
$$
D_{\x, \t}G_2(\x, \t| \x', \t')  = (1 + e^{-2\x}) \big( \delta (\t -
\t') [ {\widehat {\cal T}} (\x, t), {\widehat {\cal T}} (\x', t)]
+ \del_\t \delta(\t - \t') [ {\widehat {\cal T}} (\x, \t), {\widehat
{\cal T}} (\x', \t) ] \big)
\eqn\sixnine$$
The commutation relation of the
${\widehat {\cal T}}$ fields can be derived from the definition
\sixeighta\
and the $\u(p,q,t)$ commutation relations.
Neglecting corrections of order $e^{-x}$ we have
\def\ttt{{\widehat {\cal T}}}
\def\ddd{{\cal D}}
$$\eqalign{
[\ttt (\x, \t), \del_t \ttt(\x' ,\t) ] &\propto [\ddd_+' \ddd_+ + \ddd_-'
\ddd_-] \del_\x^2 \delta (\x - \x') \cr
[\ttt (\x, \t), \ttt(\x' ,\t) ] &\propto [\ddd_+' \ddd_+ - \ddd_-'
\ddd_-] \del_\x \delta (\x - \x') \cr}
\eqn\sixten $$
where $\ddd_\pm$ are as defined in \threefivehaa\ (the primes refer to
$\x'$).
In the limit of extremely large $\x$,  the operators $\ddd_\pm$ go as
$(\del_\x)^{-1}$ and we recover canonical communication relations. This is
essentially because in this limit the field $\ttt(\x,\t)$ becomes the same
as the ``tachyon'' $\eta$ of the standard $c=1$ matrix model (see Eq.
\threefivehb).

The non-identification of the two-point function with the propagator is
basically related to the fact that $\ddelt(u,v)$ or $\ttt(\x,\t)$ cannot
create particle states because they do not commute at equal times.

$\bullet$ How does one address the issue of propagation then?

As we have stressed, $\T(\x, \t)$ can be regarded as a Heisenberg
operator in a two-dimensional field theory.
If $H$ is a functional of $\T(\x,0)$ and $\del_\t \T(\x,0)$, then
given $\bra \psi | \T(\x,0) | \psi \ket$ and
$\bra \psi | \del_\t\T(\x,0) | \psi \ket$, one can determine in
principle $\bra \psi | \T(\x, \t) | \psi \ket $. In general it is a
difficult question whether $H$ is a functional of only $\T(\x,0),
\del_\t \T(\x,0)$. However
even if it is not clear how much initial data is
required to get a unique dynamical trajetory (unique answer for, let's
say, $\bra \psi | \T(\x, \t) | \psi \ket $), the fermionic
construction {\bf does} list {\bf all} dynamical trajectories.
Also, we do know from the analysis of small fluctuations
(for instance using the language of $\etab_\pm$) that data worth two real
functions (e.g. $\etab_\pm(\tau, t=0)$ are enough to determine the future
evolution for the fermionic state, except perhaps some discrete data
corresponding to ``discrete states''.
Incidentally, it is also clear from the
one-to-one correspondence between $\etab_\pm(\tau)$ and $\T(\x,0),
\del_\t \T(\x,0)$ that we can choose {\bf any} kind of initial data
on any given spacelike (or lightlike) surface $\t = \t_0$ (this
list includes the white hole horizon) by simply choosing the appropriate
fermionic state, or equivalently the appropriate values for
$\etab_\pm(\tau,\t_0)$.

\section {\bf Analytically continued Fermion Theory and The Euclidian
Black Hole}

In this section we
would like to study the implication for the tachyon theory of the
analytic continuation of the Fermion field theory discussed earlier in
[\DDMW], namely\foot
{This is equivalent to the analytic continuation
\REF\GROSSMILJK{D.J. Gross and N. Miljkovich, Ref. 4.}
[\GROSSMILJK] of the harmonic oscillator frequency $w \to iw$ in the
fermion potential.}
$$ t \to it, \; p \to -ip \eqn\sevenone$$
In this analytic continuation,
the ``hyperbolic transform'' becomes an ``elliptic transform'',
$$ \phi(p,q,t) = \int { dp' dq' \over [(p - p')^2 + (q - q')^2 ]^{1/2} }
\u (p,q,t), \eqn \seventwo $$
Then, by the analytically continued equation of motion,
$$  (\del_t + p \del_q - q \del_p) \u(p,q,t) = 0 \eqn\seventhree $$
we can show that $\phi (p,q,t)$ is of the form
$$ \phi (p,q,t) = T(u, \bar u)       \eqn\sevenfour $$
where
$$ u= { q - ip \over 2} \exp( -it),\; \bar u = {q + ip \over 2} \exp (it)
\eqn\sevenfive $$
As in Sec. 2,  one can show that in the low energy
approximation $T(u, \bar u)$ satisfies the equation of motion of a
massless field in the Euclidian black hole:
$$ [\barDuv ] T(u, \bar u) = 0 + o(T^2)  \eqn\sevensix $$
This corresponds to a dilaton-metric background that describes the
Euclidian black hole (the ``cigar'').

It is remarkable that the analytic continuation
\sevenone\ of the matrix model
defines the usual analytical continuation of the black hole physics!
This makes it tempting to believe that the thermal Green's functions of
the latter
may have a direct significance in terms of the matrix model in the
forbidden region.

\section{\bf Concluding Remarks}

In this concluding section we would like to comment on some issues that
need deeper understanding.

Firstly there is the question of the $S$-matrix. The definition of the
$S$-matrix requires the specification of the `in' and `out' states. These
states can be inferred by analyzing the two-point correlation function of
a complete commuting (equal time) set of field operators. A natural set is
the density operator $\psi^\dagger (x, t)\psi (x,t) \equiv \del_x
\varphi$, because perturbatively we know that $\varphi$ creates `in' and
`out' massless particle states. Once this is done we can evolve the `in'
states to the `out' states in the standard fashion and define the
$S$-matrix. This has been previously calculated
[\MSWTWO,\POLTWO,\MOORE,\DJR]
The black hole interpretation of this theory and in particular
the non-linear differential equation \threenine\ seem to strongly suggest
that the kernel that propagates `in' states to the `out' states has a
representation in the $T(u,v)$ theory. It would be very interesting to see
this explicitly.

One of the limitations of the $c=1$ matrix model as a model of black holes
is that this black hole is eternal and one cannot envisage any process
({\it e.g} formation and evaporation) that involves changing the mass of
the black hole. The simple reason  for this is that the mass of the black
hole is equal to the fermi level which cannot be changed if the number of
fermions is held fixed. It is an interesting question whether there is a
way of circumventing this difficulty within the context of string theory.

{\bf Note Added:} While this paper was being written up, we received the
article by S.R. Das,
``Matrix models and nonperturbative string propagation in two-dimensional
black hole backgrounds", Enrico Fermi Institute preprint EFI-93-16. In
this paper the author has raised the issue of the identification of the
``correct tachyon operator'' which gives rise to ``physical scattering
processes'' from a black hole. We would like to point out here that none
of the ``tachyon'' operators defined in Eq. (5) of that paper can
create particle states because they do not
commute with each other at spacelike distances since they are built out of
the density operator at different `matrix model times'. This reason is
analogous to the reason why the operator $\T(u,v)$
defined in [\DMWBH] and the present paper cannot, as has been explained in
great detail in Sec. 6. For this reason, one cannot interpret the object
defined in Eq. (12) of that paper as the wavefunction sought by the
author. We would also like to point out here that given the definitions in
Eq. (5) of that paper, it is not clear to us how time-ordered (in some
definition of time in the $u,v$-space) correlators of these operators are
related to those of the collective field theory.  For this reason the
l.h.s. of Eq. (12) of that paper cannot be obtained simply by inserting
the result (19) in the r.h.s. of (12). This needs some understanding of
the connection between (some definition of) time in the $u,v$-space and
the matrix model time. Our definition of the tachyon operator has afforded
us an immediate connection between these two ``times''. In fact, as
explained in this paper, we can use the $t$-independence in the right hand
side  of our Eq. \twooneone\ so that the matrix model time and the
``time'' of the $u,v$-space can be identified with each other and thus
relate the time-ordered correlators of $\ddelt(u,v)$ and those of the
matrix model.

\refout
\end